\documentclass[journal]{IEEEtran}
 
\usepackage{graphicx}
\usepackage{placeins}
\usepackage{float}
\usepackage{caption}
\usepackage{subcaption}
\usepackage{xcolor}
\usepackage[export]{adjustbox}
\usepackage{array}
\usepackage{booktabs}
\usepackage{amsmath,amsfonts,amssymb}
\usepackage{tabularx}
\usepackage{multirow}
\usepackage{ragged2e}
\usepackage{verbatim}
\usepackage{threeparttable}
\usepackage[acronym,shortcuts]{glossaries}
\usepackage{tablefootnote}
\usepackage{textcomp}
\newcommand{\mytexttilde}{\raisebox{0.5ex}{\texttildelow}}

\newacronym{MIMO}{MIMO}{multiple-input multiple-output}
\newacronym{D-MIMO}{D-MIMO}{distributed MIMO}
\newacronym{C-MIMO}{C-MIMO}{co-located MIMO}
\newacronym{RA-MIMO}{RA-MIMO}{repeater-assisted massive MIMO}
\newacronym{TDD}{TDD}{time-division duplex}
\newacronym{AP}{AP}{access point}
\newacronym{LoS}{LoS}{line-of-sight}
\newacronym{RIS}{RIS}{reflective intelligent surface}
\newacronym{UE}{UE}{user equipment}
\newacronym{3GPP}{3GPP}{3rd generation partnership project}
\newacronym{LTI}{LTI}{linear time-invariant}
\newacronym{AIC}{AIC}{antenna-interface circuit}
\newacronym{LNA}{LNA}{low-noise amplifier}
\newacronym{PA}{PA}{power amplifier}
\newacronym{LO}{LO}{local oscillator}
\newacronym{EVM}{EVM}{error vector magnitude}
\newacronym{ACLR}{ACLR}{adjacent channel leakage ratio}
\newacronym{SINR}{SINR}{signal-to-interference-plus-noise ratio}
\newacronym{SNR}{SNR}{signal-to-noise ratio}
\newacronym{CDF}{CDF}{cumulative distribution function}
\newacronym{ACRR}{ACRR}{adjacent channel rejection ratio}
\newacronym{NCR}{NCR}{network-controlled repeater}
\newacronym{CSI}{CSI}{channel state information}
\newacronym{MMSE}{MMSE}{minimum mean square error}
\newacronym{RF}{RF}{radio frequency}
\newacronym{NF}{NF}{noise figure}

\begin{document}

\bstctlcite{IEEEexample:BSTcontrol}

\title{Achieving Distributed MIMO Performance with Repeater-Assisted Cellular Massive MIMO}

\author{Sara Willhammar, Hiroki Iimori, Joao Vieira, Lars Sundstr\"om, Fredrik Tufvesson, and Erik G. Larsson
\thanks{Sara Willhammar and Fredrik Tufvesson are with the Department
of Electrical and Information Technology, Lund University, Sweden. Hiroki Iimori is with Ericsson Japan. Joao Vieira and Lars Sundstr\"om are with Ericsson Sweden. 
Erik G. Larsson is with the Department of Electrical Engineering (ISY), Link\"oping University, Sweden. }}

\maketitle

\begin{abstract}
In what ways could cellular massive MIMO be improved? This technology
has already been shown to bring huge performance gains. However, coverage holes
and difficulties to transmit multiple streams to multi-antenna users
because of insufficient channel rank remain issues.  Distributed
MIMO, also known as cell-free massive MIMO, might be the ultimate
solution. However, while being a powerful technology, it is expensive to install
backhaul, and it is a difficult problem to achieve accurate phase alignment for
coherent multi-user beamforming on downlink. Another option is reflective intelligent surfaces -- but they have large form factors and require a lot of training and control
overhead, and probably, in practice, some form of active filtering to
make them sufficiently band-selective.

We propose a new approach to densification of cellular systems,
envisioning \textit{repeater-assisted cellular massive MIMO}, where a
large numbers of physically small and cheap wireless repeaters are
deployed. They receive and retransmit signals instantaneously, appearing as 
\emph{active scatterers}. Meaning that they appear as ordinary channel scatterers but with amplification. We elaborate on the requirements of such repeaters, show that the performance of these systems could potentially approach that of distributed MIMO, and outline future research directions.
\end{abstract}
 
\glsresetall

\section{Introduction}
Massive \ac{MIMO} has become \emph{the} 5G technology and will
continue to be one of the cornerstones in cellular systems. Sub-6 GHz
\ac{TDD} multi-user massive \ac{MIMO} systems with reciprocity-based
beamforming can serve many terminals in the same time-frequency
resource, also in high mobility scenarios. By phase-coherent
combining, an array gain is achieved, which can significantly improve signal quality and link budget.

Despite this, there will be coverage holes.  In addition, with
multiple antenna terminals, it can be a challenge to offer more than a
single (or a few) data streams per terminal if there are not enough
independent scatters that bring up the rank of the channel.
\Ac{D-MIMO} has emerged as a promising solution to address this. In
\ac{D-MIMO}, \acp{AP} are densely deployed over a large area and due
to shorter distances between \acp{AP} and terminals, the path loss
decreases and coverage improves. Such deployment is not to be confused
with the \ac{D-MIMO} deployment currently considered in 3GPP Rel.~19
standardization, which performs coherent joint transmission between
several macro base stations. Moreover, with \ac{D-MIMO}, shadow fading
becomes less problematic due to improved macro-diversity, and
favorable propagation can be experienced even in \ac{LoS}.

Although the opportunities with \ac{D-MIMO} systems are many, they
impose deployment issues. Many of the current solutions are not standardized and/or not yet sufficient in order to get the envisioned performance gains. Questions, such as how to get the
data to the distributed units, where to perform the processing, how to
synchronize the \acp{AP}, and how to distribute the clocks and phase
references still need further development. Furthermore, the main concepts in the
standards are designed for \ac{C-MIMO} and going to \ac{D-MIMO} would
be a big step for industry to adopt. There would be a need for radical
network architectural changes and a large increase in site and
maintenance costs. In industry there can be some reluctance to big
changes and smaller steps, each beneficial, would be more likely to be
implemented.

To make the transition smoother and challenge the existing views on
how densification of current cellular systems could be done in today's sub-6~GHz systems, we
envision \textit{\ac{RA-MIMO}}.  A similar vision was presented in
\cite{compromised-reciprocity}, but extra delay, increased delay
spread and compromised channel reciprocity were identified as
obstacles; these issues are solved in our vision.  Repeaters have
traditionally been deployed to improve coverage. Coverage improvements
and capacity scaling in \ac{MIMO} systems aided by repeaters was
studied in \ac{LoS} environments~\cite{ra-mimo-los} and full-duplex
radio relays were investigated to improve outdoor-to-indoor
communication~\cite{out-in-relays}.  Today modern massive \ac{MIMO} base
stations enable multi-user \ac{MIMO} operation. Challenging the current views on how densification could be done; we propose to rethink
repeaters, could they be deployed inside the cell to support massive
\ac{C-MIMO} systems to increase macro diversity, acting as active scatterers, i.e. channel scatterers but with amplification, and potentially approach \ac{D-MIMO} performance?

The use of the terms \emph{relay} and \emph{repeater} in literature is not always consistent. 
A repeater is here considered to be the same thing as a full-duplex relay, and  \emph{instantaneously} amplifies and
re-transmits the signal, with a delay of at most a few hundred
nanoseconds. 
However,  a repeater is different from  a half-duplex relay; the difference lies in  the delay introduced. 
A half-duplex relay receives a signal of substantially longer
duration (e.g., half a slot) and then either amplifies and forwards,
or decodes and forwards, that signal. 
Repeaters (full-duplex relays) act as an active scatterer, but half-duplex relays do not.

\begin{figure*}
    \centering
    \includegraphics[width=1\linewidth, frame]{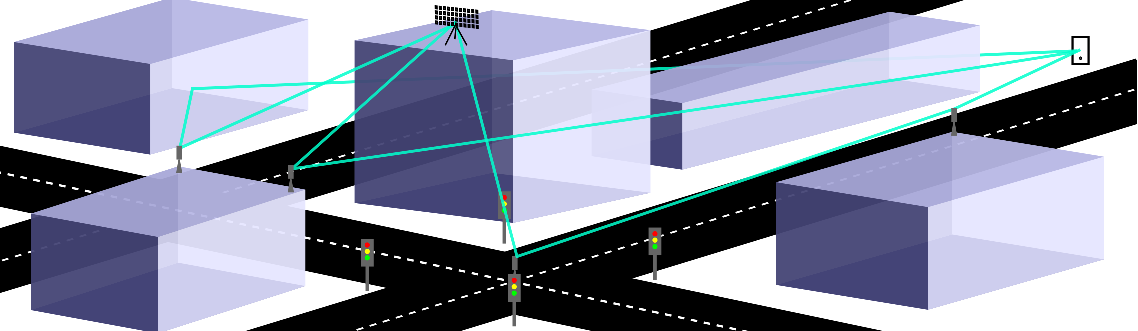}
    \caption{Envisioning repeater-assisted cellular massive \ac{MIMO}:  a massive \ac{MIMO} base station is installed on the rooftop, and assisted by multiple repeaters. }
    \label{fig:model}
\end{figure*}

Other ways to enhance performance in sub-6~GHz systems, without creating extra cells, are
to add extra \acp{AP} or, use \acp{RIS}.  Table~I presents a
comparison between these three options; these technologies are of different maturity in terms of practical implementations. A \ac{RIS} could perform the work
of a repeater while bringing benefits stemming from the many unit cells.  \Acp{NCR} are a part of Release 18~\cite{ncr-ris-5g, ris-challenges}, while \acp{RIS}  has been decided to not become a study-item;  several challenges  need to be addressed before they can be used in practice~\cite{ris-challenges}. 

Due to the different maturity of the technologies and the various possibilities to implement them, the power consumption numbers given are merely indicative. While sub-6 GHz low-power cellular repeater products on the market (not network-controlled) may consume as little as 10~W, an \ac{AP} will consume significantly more due to the required processing capabilities and backhaul interface. The low power consumption potential of \ac{RIS} yields a number at least an order of magnitude less than that of repeaters and \acp{AP}. With  \ac{RF} switch-based \ac{RIS} elements, the power consumption will be dominated by the \ac{RIS} control and switch driver circuitry~\cite{ris-power}, which may be further optimized and tailored for the application.

The advantages of deploying repeaters or
\acp{RIS}, compared to extra \acp{AP}, are that neither phase
synchronization nor backhaul is needed.
While having these benefits in common, \acp{RIS} have however drawbacks in comparison repeaters: it has a much larger form factor, requires significant control signaling, and incurs
significant training and configuration overhead.  Furthermore, a passive, or semi-passive \ac{RIS} would not be nearly as band-selective as a repeater, which can be frequency-selective down to the level of a single channel.  
An active \ac{RIS}, which could amplify signals, could act
similar as a repeater but due to regulatory requirements it would
likely require a large number of expensive components to filter the
signal.  As a first-order approximation, to achieve the same
signal amplification as a repeater, a [passive] \ac{RIS} would require
a number of unit cells equal to the square root of the repeater
(power) gain. To replace a 60-dB gain repeater, a \ac{RIS} would need
in the order of 31~\texttimes~31~\mytexttilde~1000 unit cells (for
simplicity, we ignore the directivity of the patch antenna patterns
here). Furthermore, to obtain the macro diversity, one would need many
distributed units, which appears challenging with \acp{RIS}. Challenges with incorporating \acp{RIS} into cellular networks were discussed in~\cite{ris-challenges}, including a simulated comparison to \acp{NCR}, showing a performance gain in terms of spectral efficiency to the latter.

\begin{table}[t]
  \caption{Repeaters versus \acp{RIS} versus extra APs for performance enhancement.}
  \label{tab:comparison}
  \centering
  \begin{threeparttable}
    \begin{tabular}{c|c|c|c}   
        \textit{\textbf{Required \textbackslash Assisted by}}             & \textbf{Repeater} & \textbf{\ac{RIS}} & \textbf{Extra \ac{AP}}\\
       \hline
       Form factor  & Small & Large  & Small \\
       \hline
       Power consumption  & \mytexttilde 10 W & \mytexttilde 1 W & \mytexttilde 20 W \\
       \hline
       Instantaneous CSI & No & Yes & Yes \\
       \hline
       Extra training & Little & Large  & Little \\
       \hline
       Phase synchronization & No & No  & Yes\\
       \hline
       Backhaul  & No & No  & Yes \\
       \hline
       Active radiated power & Yes & No  & Yes\\
       \hline
       Noise source & Yes & No & Yes \\
       \hline
       Array gain  & Some & High & Some \\
       \hline
       Control overhead & Low & High & None\\
       \hline
       Deployment cost & Low & Medium & High\\
       \hline
       Deployment flexibility & High & Low & Medium\\
       \hline
       Band-selectivity & Yes & Little & Yes\\
       \hline
    \end{tabular}
    \end{threeparttable}
\end{table}

\section{Envisioning repeater-assisted massive MIMO}

Within a cell covered by a massive \ac{MIMO} base station, the
deployment of multiple repeaters could be simple and inexpensive: they
have a small form factor and only require a power supply.  The massive
\ac{MIMO} base station will remain in control of all multi-antenna
signal processing, enabling full backward compatibility and requiring
no major system upgrades. The base station will control the repeaters
and configure their parameters (see \cite{Gang15} for possible
configurations) via, e.g., a control channel. No significant control
signaling is required and the NCR 3GPP framework could be sufficient
for the purpose, with the addition of power
control. Fig.~\ref{fig:model} shows a deployment example. The massive
\ac{MIMO} array is placed on the rooftop and the terminal is a
handheld device. Four repeaters are deployed: three on rooftops and
one on a traffic light.  Repeaters could be deployed in a wide variety
of ways: on separate masts, on building facades or rooftops, on lamp
posts and traffic lights, to mention some possibilities.  An advantage
is that the deployment does not need to be too carefully planned;
repeaters can be installed in a ``plug-and-play'' fashion. The
coverage zones of different repeaters can overlap and they can pick up
signals from each other, as long as positive feedback effects are
under control \cite{larsson2024stability}.

The signal received at the base station or the terminal will be a
superposition of signals traveling via direct paths, including
\ac{LoS} and/or multipath reflections, and signals going via
repeater(s). This means that the channels between a terminal and the
base station is a result of several components: 1) the direct path
between the terminal and base station, 2) paths between a terminal and
a repeater, 3) paths between a repeater and a scatterer, 4) paths
between repeaters, 5) paths between a repeater and the base station,
and 6) paths between a scatterer and the base station. As this is only
concatenations of \ac{LTI} systems, the result is also an \ac{LTI}
system. Its response is (in principle) the same in uplink and
downlink, enabling reciprocity-based multi-user beamforming. Repeaters
should act as active scatterers in the environment, i.e. similar to ordinary channel scatterers, but with higher gain. The
repeaters will naturally introduce a delay, i.e. a frequency-dependent
phase shift, which is typically not a problem as it is comparable to
the frequency response of a far-away scatterer.

\section{Repeaters as active channel scatterers}

\subsection{Requirements on the repeater}

To support the vision of \ac{RA-MIMO} systems, the key aspect is that it must be possible to realize repeaters that have the
same properties as channel scatterers present in the
propagation environment. With the same properties, they would be a transparent part from both
the base station and terminal perspective. To act as a channel
scatterer, the repeater must satisfy the following requirements:

\begin{enumerate}
    \item The repeater should \emph{\textbf{have an \ac{LTI}
      response}} and hence not introduce non-linearities and
      out-of-band radiation. It should not have an automatic gain
      control circuit that introduces fast time-variations of the
      amplification gain.
      
    \item For reciprocity-based transmission, the contribution of the
      repeater should be the same in uplink and downlink, i.e. the
      repeater should  \emph{\textbf{be reciprocal}}. The forward gain
      of the repeater should, for each subcarrier, be equal to its
      reverse gain.
      
    \item The excess delay of the channel collectively comprised by
      the ordinary channel scatterers and the repeaters must be
      smaller than the duration of the cyclic prefix. Hence, each
      repeater may only \emph{\textbf{introduce \emph{a much smaller}
        delay than the duration of the cyclic prefix}}, which requires
      an almost instantaneous processing and amplification.
      
    \item The repeater should \emph{\textbf{be band-selective}}, i.e.,
      only amplify signals within the desired band. This requires the
      use of sharp filters, which have long impulse responses -- but
      whose length still fits within the duration of the cyclic
      prefix.
      
    \item Noise is unavoidable and stems from multiple sources:
      foremost thermal noise and phase noise. There is a
      limit on how noisy a repeater can be in order to
      \textbf{\textit{not act as a limiting noise source}}.
      
    \item Repeaters can be categorized into single-antenna and
      dual-antenna repeaters; for the latter, the repeater
      implementation should \emph{\textbf{align with the TDD
        pattern}}.
  \end{enumerate}
    
\subsection{Single-antenna repeaters}

\begin{figure}
    \centering
    \includegraphics[scale=0.3]{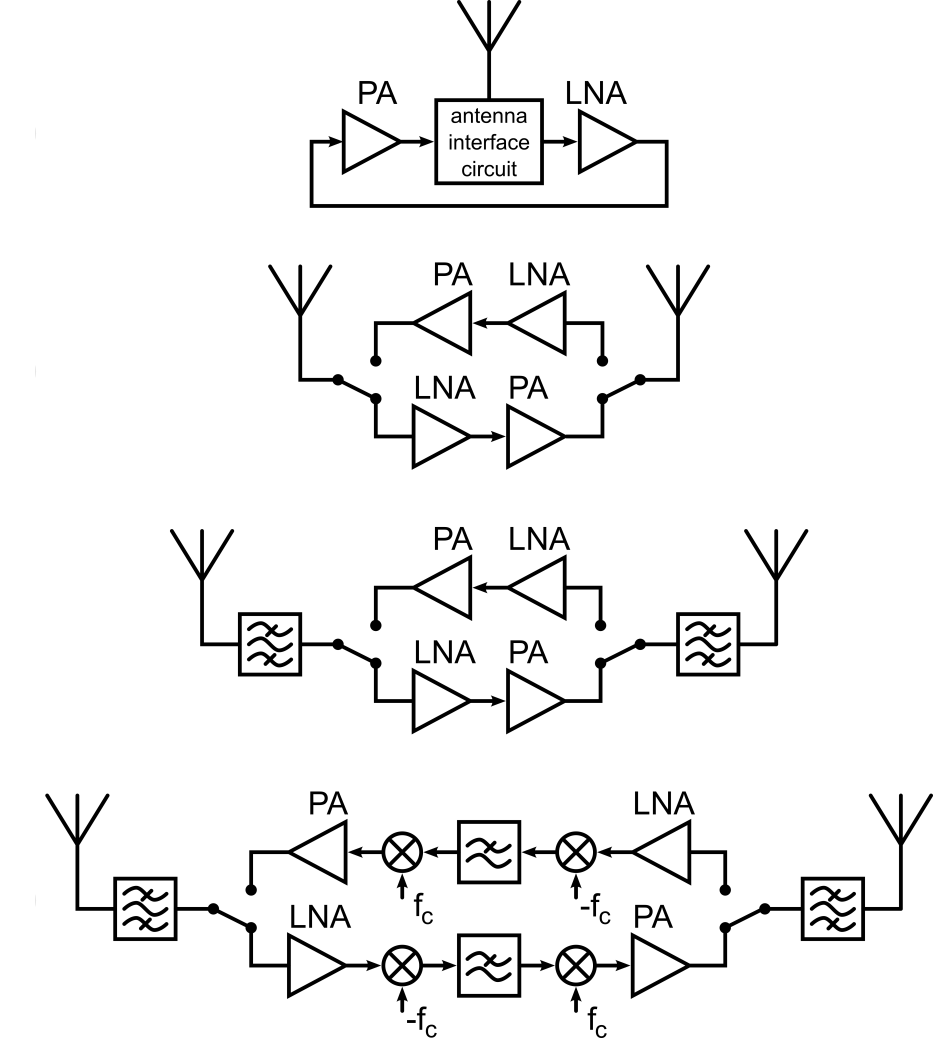}
    \caption{Principle of 1) single-antenna and 2) dual-antenna repeater architectures, 3) with RF filters, and 4) with up- and down-conversion for baseband filtering.}
    \label{fig:all_repeater_arch}
\end{figure}

Repeaters only amplify the received signals, without performing any
processing. In a \emph{single-antenna repeater}, the same antenna is
used simultaneously for reception and transmission and is shared
between the input and output of the amplifier chain; see 1) in
Fig.~\ref{fig:all_repeater_arch}.  This antenna is connected to an
\ac{AIC} that forwards the received signal to a \ac{LNA}, which in
turn forwards the signal to a \ac{PA} that generates the transmit
signal. The \ac{AIC} also forwards the \ac{PA} output to the antenna.
Note that the physical implementation of the antenna is not restricted
to a single \emph{physical} antenna, it could be an antenna array with
beamforming capabilities.

Important \ac{AIC} parameters are the level of \ac{PA}-to-antenna and antenna-to-\ac{LNA} losses, and the \ac{PA}-to-\ac{LNA} isolation. The latter dictates the maximum gain of the repeater: the gain   must be lower than the isolation to avoid self-oscillation. 
The antenna-to-\ac{LNA} loss adds to the repeater's \ac{NF} and the \ac{PA}-to-antenna loss reduces the transmit  power.

The \ac{AIC} may be implemented as a circulator or as an electrical-balance duplexer. \ac{PA}-to-\ac{LNA} isolations of 20--50 dB have been reported, with the highest levels only achieved with a relative bandwidth of up to 10\% \cite{Chen2021}. Assuming a 50~dB isolation, the single-antenna repeater may only allow for a gain up to some 40 dB (assuming a 10 dB margin to avoid self-oscillation). The \ac{PA}-to-antenna loss introduced by the \ac{AIC} is in most cases within 2--4~dB for the sub-6 GHz range and the \ac{NF} contribution is within 2.5--6~dB \cite{Chen2021}. 

While having a limited gain potential, the single-antenna repeater   intrinsically supports reciprocity-based communication.

\subsection{Dual-antenna repeaters}

The repeater architecture considered in recent standardization efforts on \ac{NCR}~\cite{3gpp-repeaters} is the dual-antenna repeater. 
Dual-antenna repeaters are  substantially more robust than   single-antenna repeaters, and much more resilient to self-oscillation.
One antenna is assigned for the base station side and another for the terminal side, see 2) in Fig.~\ref{fig:all_repeater_arch}, with one amplification path for the uplink and another for the downlink. A switch selects the transmission direction. 
Dual-antenna   repeaters naturally provide high isolation between the \ac{PA} output and the \ac{LNA}, as the two antennas can be physically designed and placed to provide a low inter-antenna gain. Therefore, dual-antennas repeaters allow for substantially higher gains, compared to  single-antenna repeaters. Furthermore, design compromises associated with the \ac{AIC} in  single-antenna repeaters are avoided. In particular, a dual-antenna repeater has a lower \ac{NF} and a higher output power capability.  

The dual-antenna repeater does however not support reciprocity-based communication unless the uplink and downlink signal paths have matching transfer functions across the band. Even with the signal paths \emph{designed} to be identical, when temperature changes, their relative gains and phases  will change. This calls for periodic reciprocity calibration \cite{larsson2024reciprocity}.

Existing repeater requirements~\cite{ncr-ris-5g} include metrics  such as \ac{EVM}, \ac{ACLR}, and emission limits. 
No maximum gain is specified for the   passband of the repeater, but there are requirements on the maximum allowed gain outside the passband, 60~dB up to 1~MHz from the passband edge, dropping to 35~dB at 10~MHz offset. 
Also,   the ratio between the passband gain and the adjacent channel gain must exceed  33~dB or 45~dB, depending on   repeater class and carrier frequency. 

\Ac{RF} band filters can be used, see 3) in Fig.~\ref{fig:all_repeater_arch}, to attenuate out-of-band signals before the \ac{LNA}, and to comply with out-of-band emission requirements. However, such filters may not be sufficient to fulfill gain requirements outside the repeater passband. Better selectivity can be achieved by  baseband filtering, as shown in 4) in Fig.~\ref{fig:all_repeater_arch}: the received signal is then down-converted to baseband,  filtered, and  up-converted using the same \ac{LO}.  

Baseband filtering may also be carried out in the digital domain, enabling the use of sharper and more flexible filters. This comes at  increased cost and complexity, due to the digital processing itself, and to the addition of analog-to-digital and digital-to-analog converters.

\subsection{Hardware performance and impairments}

\subsubsection{Additive noise}
In the architecture in 3) in Fig.~\ref{fig:all_repeater_arch}, both the \ac{RF} band filter  and the antenna-to-\ac{LNA} switch introduce losses and thus an increase in the repeater's \ac{NF}. The filter loss varies greatly depending on the type of filter technology and the target band. A typical terminal band filter at sub-6 GHz frequencies has a loss of around 2~dB while the switch loss is only a fraction of a dB.

The \ac{NF} for the remaining part of the signal chain is dominated by the \ac{LNA}.  
CMOS \acp{LNA} operating in the sub-6 GHz range have a \ac{NF} of 2--3~dB  \cite{Belostotski20}. 
Adding a 2~dB  filter loss yields a total  \ac{NF} of up to 5~dB.

\subsubsection{Amplification, distortion, and power}
The \ac{LNA} may provide a gain of some 20--30~dB for sub-6 GHz frequencies. With additional amplifiers  prior to the final \ac{PA},  a total gain of 60~dB and beyond can be achieved. As the  amplification level  increases, so does the non-linear distortion generated by the amplifiers. The overall distortion  should be dominated by the \ac{PA} as there is most to gain in  power consumption by optimizing the \ac{PA} efficiency. 

A fully analog repeater can not use digital linearization techniques to reduce the \ac{PA} distortion. However, limited improvements may be achieved using analog linearization techniques. Backing off the output power reduces distortion, but  comes at the expense of  decreased efficiency. The required backoff depends on  both the targeted \ac{ACLR} level and the \ac{PA}'s non-linearity characteristics. An estimate may be calculated assuming  a third-order polynomial  \ac{PA} model and   an OFDM input signal. 
The output power for a desired \ac{ACLR} is then proportional to the \ac{PA}'s 1-dB gain compression point, and inversely proportional to the square root of the \ac{ACLR}~\cite{3gpp-7-24}. 

The \ac{PA}'s power capability depends on both semiconductor technology and frequency \cite{WangPASurvey}. For low-cost repeaters, targeting ``local area class'' specifications and operating in the sub-6~GHz range, it is reasonable to use terminal-grade components. Therefore, the output power target  should be in the 20~dBm regime (comparable to  cellular and WiFi terminals). 
Repeaters with higher output power are subject to  ``wide area class'' specifications and must fulfill  more stringent emission requirements. 
The \ac{ACLR} requirements for wide-area and local-area repeaters are 45~dB and 31~dB, respectively~\cite{3gpp-repeaters}. 
 
\subsubsection{Phase noise}
Repeaters with baseband filtering   use an \ac{LO} signal for up- and down-conversion, and this  \ac{LO} signal exhibits phase noise.
In conventional transceivers, phase noise  leads to \ac{EVM} degradation and increased emissions close to the transmitted carrier. However, in   baseband-filtering repeaters, the same \ac{LO} signal may be used for both up- and down-conversion. This significantly reduces the impact of the phase noise as the phase noise introduced in the down-conversion will be canceled in the up-conversion. However, the level of cancellation depends on the delay of the signal introduced between the up- and down-conversion \cite{Riihonen2012_Phase_noise}. 
This delay is dominated by the group delay of the baseband filter, and with filters having substantially larger bandwidth than the phase noise spectrum of the \ac{LO}, the impact of phase noise will be low. Similarly, \ac{LO} phase noise will not cause any phase drift of the repeater.

\subsubsection{I/Q impairments}
To down-convert a signal to baseband (and back to \ac{RF}), quadrature mixers are used to generate complex baseband equivalents of the \ac{RF} signal. Quadrature mixers also have impairments, namely I/Q gain and phase mismatch as well as DC offset. If left uncorrected, an I/Q mismatch will degrade the \ac{EVM} and a DC offset will lead to a transmitted signal with LO signal leakage. In sub-6 GHz bands, I/Q gain and phase errors around 1\% and 1~degree, respectively, are typically achieved without correction. The same errors in both the up- and down-conversion result in an \ac{EVM} contribution of 2\%.

\subsubsection{Delay}
A repeater introduces delay to the signal. Both the \ac{RF} and baseband filters,   and the amplifiers, contribute to this delay.
A representative example is an analog 5th-order lowpass Butterworth filter with 10~MHz bandwidth that has a group delay of some 50~ns, scaling inversely proportionally to the filter bandwidth. For digital filtering and associated conversion between analog and digital domains, the processing delay may reach 100's of ns. That is only a small fraction of the cyclic prefix and therefore insignificant. 

\section{Performance of repeater-assisted cellular massive MIMO}

After elaborating on how to implement repeaters to act as active scatterers we here aim to show the potential of \ac{RA-MIMO}, comparing it to massive
\ac{C-MIMO}, small cells, and densely deployed \ac{D-MIMO}, which serves as a
benchmark for ideal performance, e.g. assuming no synchronization
issues or other impairments.  We consider multi-user \ac{MIMO}
transmission, where either a massive \ac{MIMO} array located at the
base station (first two cases) or distributed \acp{AP} (last two cases)
receive uplink data from multiple terminals, whereas the repeaters in \ac{RA-MIMO} are
located at the same positions as the \acp{AP} in the last two cases.  For simplicity, the results disregard interactions between repeaters, however, the maximum amplification gain is chosen such that the system becomes stable without positive feedback (using the criteria in \cite{larsson2024stability}). The \ac{SINR} after the uplink combiner is used as performance metric, with conventional \ac{MMSE} combining and perfect \ac{CSI} assumed unless otherwise stated.
{\Acp{CDF} based on both channel distributions and user locations are generated, using 100,000  realizations. The terminals are randomly positioned in each realization.
For the \ac{AP} selection in the small cell scenario, a greedy strategy leveraging the Hungarian algorithm to optimize total signal strength across all terminals is used.

\begin{figure}[t!]
    \centering
    \includegraphics[width=\linewidth]{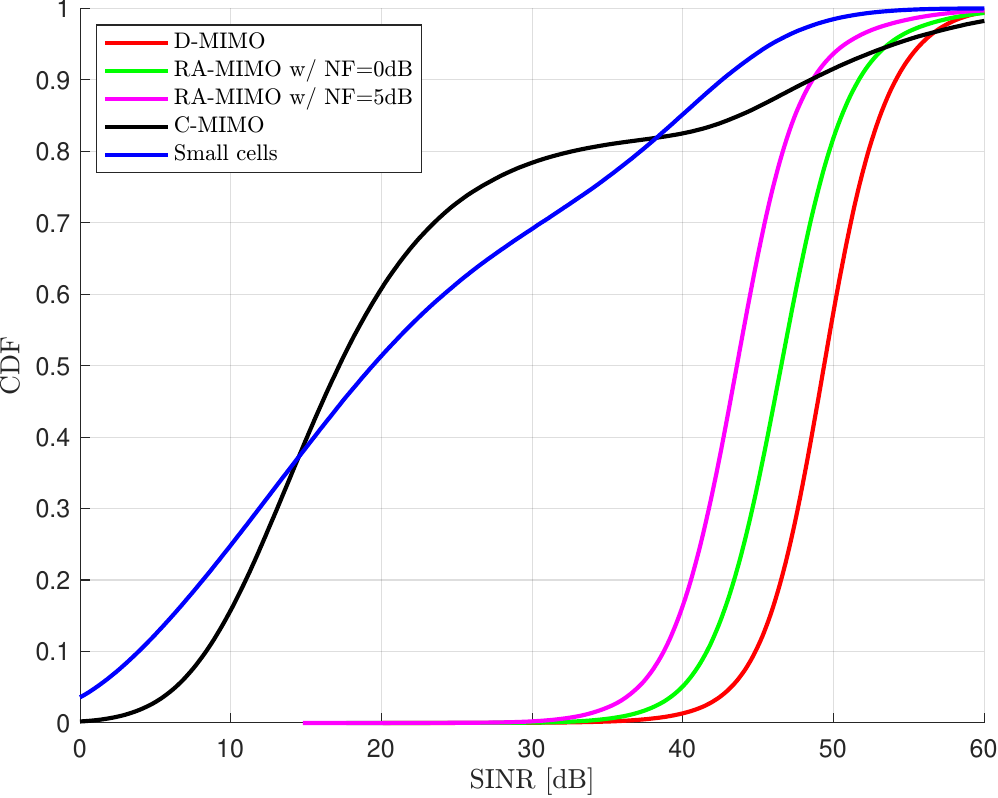}
       \caption{Simulated SINR distributions for all terminals, for \ac{D-MIMO}, \ac{C-MIMO}, ideal-hardware \ac{RA-MIMO} (0 dB \ac{NF}), realistic \ac{RA-MIMO} (5~dB \ac{NF}), and small cells.}
    \label{fig:sim:comp_diff_arch}
    \vspace{1ex}
    \includegraphics[width=\linewidth]{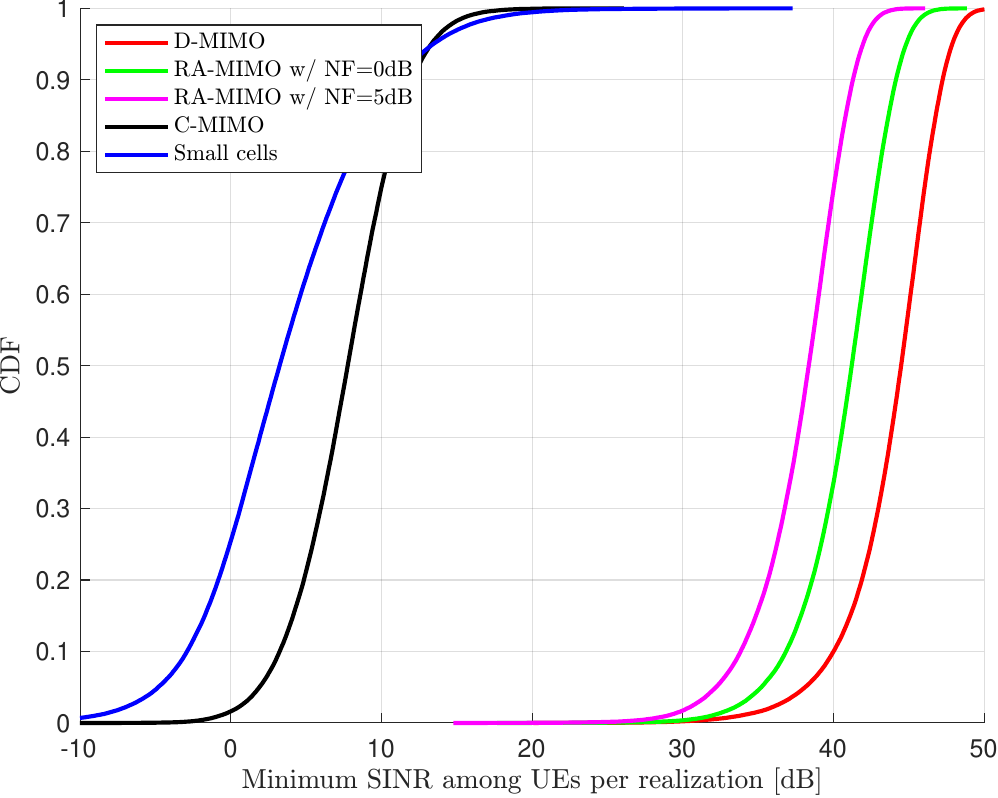}
       \caption{Simulated distributions of the worst \ac{SINR} among all terminals per realization, for the same
       systems as in Fig.~\ref{fig:sim:comp_diff_arch}.}
    \label{fig:sim:worst_sinr}
\end{figure}

The simulation is conducted in an outdoor square area with 400~m
sides. At the center, the base station with a 64-antenna uniform
circular array is located at a 20~m height. The carrier frequency is
3.6~GHz with 20~MHz bandwidth. The noise temperature is 290~Kelvin and the
\ac{NF} is 5~dB. Path loss and \ac{LoS} probability models stem from
3GPP UMi TS 36.814. The fading is Ricean with a 10~dB K-factor if
\ac{LoS} exists, otherwise it is Rayleigh. The 64 single-antenna
repeaters/\acp{AP} are distributed on a uniform grid at a
10~m height above the terminals. The terminals and repeaters have
maximum 20~dBm output power. In addition, the repeaters are regulated by a maximum 70~dB amplification
gain. The actual gain used at each repeater is determined as the
maximum allowable amplification gain under these two constraints. To
demonstrate the potential performance of RA-MIMO, it is assumed that
the repeater-base station channels follow Rician fading with a
\ac{LoS} component. In practice, this can not always be guaranteed but is still likely with some deployment considerations.

Considering the additive noise at the base station and the repeaters,
we note that the combination of 1) received signal power
variations across repeaters with 2) additive noise power at the
repeaters may result in low-quality repeated signals that may
deteriorate system performance (i.e., some repeaters act as noise
sources). We address these challenges with a modified \ac{MMSE}
combiner to incorporate the amplified repeater noise stemming from
repeater-base station channels into the matrix inversion of the linear
\ac{MMSE} filtering, in case of RA-MIMO.

Fig. \ref{fig:sim:comp_diff_arch} compares simulated \acp{CDF} of the
\ac{SINR} for different systems, where the two \ac{MIMO} architectures
are compared to \ac{RA-MIMO} systems with a 70~dB maximum
amplification gain at each repeater.  The performance of \ac{RA-MIMO}
systems with a 0~dB \ac{NF} at each repeater is included as reference
to depict an optimistic scenario with ideal-hardware repeaters.  This
enhancement in \ac{SINR} signifies that \ac{RA-MIMO} brings diversity
and richer communication channels, thereby improving the
low-performing terminals and bridging the performance gap between
\ac{C-MIMO} and \ac{D-MIMO} systems. \ac{RA-MIMO} systems can offer substantial benefits, but certain
operational considerations must be taken into account. For instance,
there is a risk for positive feedback loops between repeaters due to
over-amplification if the amplificaiton control is not carefully
designed, which can introduce instability in the system
\cite{larsson2024stability}. Additionally, the amplification process
may also amplify noise, affecting the signal
quality. These factors should be carefully managed in practice to
fully realize the advantages of RA-MIMO.

Shifting our focus 
to the worst-case \ac{SINR} performance among terminals for each realization, this analysis provides insights into the robustness and reliability of the \ac{MIMO} configurations under challenging conditions.  
Fig. \ref{fig:sim:worst_sinr} presents the \acp{CDF} of the worst \ac{SINR} among terminals for the different architectures. As depicted, \ac{RA-MIMO} systems exhibit significant improvements over \ac{C-MIMO} systems in terms of the worst communication performance, approaching the performance levels of \ac{D-MIMO}. Together with the findings in Fig.~\ref{fig:sim:comp_diff_arch}, these results indicate that \ac{RA-MIMO} systems have the potential to achieve a level of \ac{SINR} performance uniformity that is comparable to that of \ac{D-MIMO}.

\section{Open questions and future work}
\noindent{\textbf{System design and adaptions:}} Although being an easy-to-deploy solution, some minor adaptions of the network operation might be needed for further performance improvements. Inter-cell interference is, e.g., not considered here.

\noindent\textbf{Repeater design:} Many of the performance metrics associated with repeaters are common with both base stations and terminals, such as linearity, noise, delays, energy consumption and frequency selectivity, which are all important research topics. One important research question regarding repeaters is how well a repeater can be made reciprocal by design and by means of calibration. Further, the best choice of  antennas and antenna patterns is also a research opportunity.

\noindent{\textbf{Multi-antenna repeaters and beamforming:}} Performance gains are possible with more antennas. For \ac{RA-MIMO}, this opens up questions such as how to beamform in presence of repeater impairments and how to compute the beamforming weights with minimal signaling overhead.

\noindent\textbf{Repeater deployment strategies:} One benefit of repeater deployments is that it does not need to be carefully planned, although optimizations are possible. Aspects to consider are the distribution of repeaters, areas in the cell in need of improved coverage, and areas with few scatterers. Channel analyses would be beneficial in this regard.

\noindent\textbf{Repeater configuration:} Some configuration of the repeaters is required and configuration strategies need to be assessed. Questions include how to jointly optimize repeater amplification and activation settings.

\noindent\textbf{Frequency-division duplex:} The focus here has been on TDD-based systems but the solution could potentially be used in frequency-division duplex-based systems. Opportunities and challenges 
regarding that are left as an open problem.

\section{Acknowledgements}
This work was supported by the strategic research area ELLIIT, the KAW foundation, the Swedish Research Council, and H2020-REINDEER.

\bibliographystyle{IEEEtran}

\bibliography{paper}

\vspace*{-10mm}

\begin{IEEEbiographynophoto}{SARA WILLHAMMAR} (Member, IEEE) has a dual Ph.D. degree from Lund University and KU Leuven. Now she is a postdoctoral fellow at Lund University, focusing on channels and propagation in massive MIMO systems.
\end{IEEEbiographynophoto}

\vspace*{-10mm}

\begin{IEEEbiographynophoto}
{HIROKI IIMORI} (Member, IEEE) earned his Ph.D. with special distinction from Jacobs University Bremen, Germany, in 2022. He is currently with Ericsson Research, focusing on signal processing and wireless communications.
\end{IEEEbiographynophoto}

\vspace*{-10mm}

\begin{IEEEbiographynophoto}{JOAO VIEIRA} received the PhD degree from Lund University, Sweden, in 2017. Since then, he is with Ericsson Research investigating different 6G candidate technologies, especially those comprising large antenna arrays.
\end{IEEEbiographynophoto}

\vspace*{-10mm}

\begin{IEEEbiographynophoto}{LARS SUNDSTR\"OM} (Senior Member, IEEE) is with Ericsson Research as a Principal Researcher with interests ranging from wideband direct-RF data converters to radio architectures for cellular systems.
\end{IEEEbiographynophoto}

\vspace*{-10mm}

\begin{IEEEbiographynophoto}{FREDRIK TUFVESSON} (Fellow, IEEE)  is a Professor at Lund University, Sweden. His main research interest is the interplay between the channel and the rest of the communication system with applications in 5G/B5G systems.
\end{IEEEbiographynophoto}

\vspace*{-10mm}

\begin{IEEEbiographynophoto}{ERIK G. LARSSON} (Fellow, IEEE)  is a Professor at Link\"oping University, Sweden.  
His interests are in  wireless communications, statistical signal processing, decentralized machine learning, and network science.

\end{IEEEbiographynophoto}

\end{document}